\documentclass{llncs}
\usepackage{llncsdoc}
\usepackage{color}
\usepackage{url}
\usepackage{amsfonts}
\usepackage{amssymb}
\usepackage{epsfig}
\usepackage{multirow}
\usepackage{graphicx}

\newcommand{\hometag}{\emph{home}}
\newcommand{\awaytag}{\emph{away}}
\begin{document}
\title{Predicting the NFL Using Twitter}
\author{Shiladitya Sinha\inst{1}\and Chris Dyer\inst{1}\and Kevin Gimpel\inst{2}\and Noah~A.~Smith\inst{1}}
\institute{Carnegie Mellon University, Pittsburgh PA 15213, USA\\
\and
Toyota Technological Institute at Chicago, Chicago IL 60637, USA}
\maketitle

\begin{abstract}
We study the relationship between social media output and National Football League (NFL) games, using a dataset containing messages from Twitter and NFL game statistics. 
Specifically, we consider tweets pertaining to specific teams and games in the NFL season and use them alongside statistical game data to build predictive models for future game outcomes (which team will win?) and sports betting outcomes (which team will win with the point spread? will the total points be over/under the line?). We experiment with several feature sets and find that simple features using large volumes of tweets can match or exceed the performance of more traditional features that use game statistics. 


\end{abstract}
\section{Introduction}

Twitter data has been used to predict and explain a variety of real-world phenomena, including opinion polls~\cite{oconnor:2010}, elections~\cite{tumasjan:2010}, the spread of contagious diseases~\cite{paul:2011}, and the stock market~\cite{bollen:2011}. This is evidence that Twitter messages in aggregate contain useful information that can be exploited with statistical methods. In this way, Twitter may offer a way to harness the ``wisdom of crowds'' \cite{surowiecki-05} for making better predictions about real-world events. 


In this paper, we consider the relationship between National Football League (NFL) games and the Twitter messages mentioning the teams involved, in order to make predictions about games. We focus on the NFL because games are broadcast widely on television throughout the US and teams play at most once per week, enabling many to comment on games via social media. 
NFL football also has active betting markets. The most well-known is the point spread line, which is a handicap for the stronger team chosen by bookmakers to yield equal bets on both sides. 
Factoring in the bookmaker's commission, a betting strategy that predicts the winner ``with the spread'' in more than 53\% of games will be profitable. 
In this paper, we build models to predict game and betting outcomes, considering a variety of feature sets that use Twitter and game statistical data. 
We find that simple features of Twitter data can match or exceed the performance of the game statistical features more traditionally used for these tasks. 

Our dataset is provided for academic research at \url{www.ark.cs.cmu.edu/football}. It is hoped that our approach and dataset may be useful for those who want to use social media to study markets, in sports betting and beyond. 

\section{Problem Domain and Related Work}
Each NFL regular season spans 17 weeks from September to January, with roughly one game played per week by each team. 
In each game, the \textbf{home team} plays at their own stadium and hosts the \textbf{away team}. The most popular wager in NFL football is to choose the team that will win given a particular handicap called the \textbf{point spread}. The point spread is a number set by bookmakers that encodes the handicap for the home team. It is added to the home team's score, and then the team with the most points is called the winner \textbf{with the spread (WTS)}. For example, if the NY Giants are hosting the NY Jets and the point spread is $-4$, then the Giants will have to win by at least $4$ in order to win WTS. If the Giants win by fewer than $4$, the Jets win WTS.\footnote{If the Giants win by \emph{exactly} $4$, the result is a \textbf{push} and neither team wins WTS.} Also popular is to wager on whether the total number of points scored in the game will be above or below the \textbf{over/under line}. 

Point spreads and over/under lines are set by sports betting agencies to reflect all publicly available information about upcoming games, including team performance and the perceived outlook of fans. 
Assuming market efficiency, one should not be able to devise a betting strategy that wins often enough to be profitable. 
In prior work, most have found the NFL point spread market to be efficient overall \cite{lacey-90,levitt04,boulier-06}, or perhaps only slightly inefficient \cite{dare-96,dare-04}. 
Others pronounced more conclusively in favor of inefficiency~\cite{zuber85inefficient,golec-91}, but were generally unable to show large biases in practice~\cite{gray-97}.\footnote{Inefficiencies have been attributed to bettors overvaluing recent success and undervaluing recent failures \cite{vergin-01}, cases in which home teams are underdogs~\cite{dare-04}, large-audience games, including Super Bowls~\cite{dare-96}, and extreme gameday temperatures~\cite{borghesi-07}.}
Regardless of efficiency, several researchers have designed models to predict game outcomes~\cite{harville80predictions,stern91probability,glickman-98,knorr00dynamic,gimpel-06,baker-13}. 

Recently, Hong and Skiena \cite{hong-10} used sentiment analysis from news and social media to design a successful NFL betting strategy. However, their main evaluation was on in-sample data, rather than forecasting. Also, they only had Twitter data from one season (2009) and therefore did not use it in their primary experiments. We use large quantities of tweets from the 2010--2012 seasons and do so in a genuine forecasting setting for both winner WTS and over/under prediction.



\section{Data Gathering}

We used Twitter (\url{www.twitter.com}) as our source of social media messages (``tweets''), using the ``garden hose'' (10\%) stream to collect tweets during the 2010--2012 seasons. For the 2012 season, this produced an average of 42M messages per day. We tokenized the tweets using \texttt{twokenize}, a freely available Twitter tokenizer developed by O'Connor et al.~\cite{oconnor10twokenize}.\footnote{\url{www.ark.cs.cmu.edu/TweetNLP}}
We obtained NFL game statistics for the 2010--2012 seasons from NFLdata.com (\url{www.nfldata.com}). The data include a comprehensive set of game statistics as well as the point spread and total points line for each game obtained from bookmakers. 

\begin{small}
\begin{table}[tb]
\begin{center}
\caption{Hashtags used to assign tweets to New York Giants (top) and New York Jets (bottom). If a tweet contained any number of hashtags corresponding to exactly one NFL team, we assigned the tweet to that team and used it for our analysis.\label{tab:hashtags}}
\begin{tabular}{p{12cm}}
{\tt \#giants} {\tt \#newyorkgiants} {\tt \#nygiants} {\tt \#nyg} {\tt \#newyorkfootballgiants} {\tt \#nygmen} {\tt \#gmen} {\tt \#gogiants} {\tt  \#gonygiants} {\tt  \#gogiantsgo} {\tt  \#letsgogiants} {\tt  \#giantsnation} {\tt  \#giantnation} \\
\hline
{\tt \#jets} {\tt  \#newyorkjets} {\tt  \#nyjets} {\tt  \#jetsjetsjets} {\tt  \#jetlife} {\tt  \#gojets} {\tt  \#gojetsgo} {\tt  \#letsgojets} {\tt  \#jetnation} {\tt  \#jetsnation} \\
\end{tabular}
\end{center}
\end{table}
\end{small}

\subsection{Finding Relevant Tweets}

Our analysis relies on finding relevant tweets and assigning them to particular games during the 2010--2012 NFL seasons. 
We can use timestamps to assign the tweets to particular weeks of the seasons, but linking them to teams is more difficult. 
We chose a simple, high-precision approach based on the presence of hashtags in tweets. We manually created a list of hashtags associated with each team, based on familiarity with the NFL and validated using search queries on Twitter. There was variation across teams; two examples are shown in Table~\ref{tab:hashtags}.\footnote{Although our hashtag list was carefully constructed, some team names are used in many sports. After noticing that many tweets with \texttt{\#giants} co-occurred with \texttt{\#kyojin}, we found that we had retrieved many tweets referring to a Japanese professional baseball team also called the Giants. So we removed tweets with characters from the Katakana, Hiragana, or Han unicode character classes.} 
We discarded tweets that contained hashtags from more than one team. We did this to focus our analysis on tweets that were comments on particular games from the perspective of one of the two teams, rather than tweets that were merely commenting on particular games without associating with a team. 
When making predictions for a game, our features only use tweets that have been assigned to the teams in those games.

For the tasks in this paper, we created several subsets of these tweets. 
We labeled a tweet as a {\bf weekly tweet} if it occurred at least 12 hours after the start of the previous game and 1 hour before the start of the upcoming game for its assigned team. {\bf Pregame tweets} occurred between 24 hours and 1 hour before the start of the upcoming game, and {\bf postgame tweets} occurred between 4 and 28 hours after the start of the previous game.\footnote{Our dataset does not have game end times, though NFL games are nearly always shorter than 4 hours. Other time thresholds led to similar results in our analysis.} 
Table~\ref{tab:vols} shows the sizes of these sets of tweets across the three NFL seasons.


\begin{small}
\begin{table}[t]
\begin{center}
\small
\caption{Yearly pregame, postgame, and weekly tweet counts.}\begin{tabular} {c | r | r | r}
season & pregame & postgame & weekly \\ \hline
2010 & 40,385 & 53,294 & 185,709 \\ 
2011 & 130,977 & 147,834 & 524,453 \\ 
2012 & 266,382 & 290,879 & 1,014,473 \\
\end{tabular}
\end{center}
\label{tab:vols}
\end{table}
\end{small}

To encourage future work, we have released our data for academic research at \url{www.ark.cs.cmu.edu/football}. It includes game data for regular season games during the 2010--2012 seasons, including 
the point spread and total points line. We also include tweet IDs for the tweets that have been assigned to each team/game. 

\section{Data Analysis}
\label{sec:data_analysis}
\label{sec:tweetclassification}
Our dataset enables study of many questions involving professional sports and social media. We briefly present one study in this section: we measure our ability to classify a postgame tweet as whether it follows a win or a loss by its assigned team. 
By using a classifier with words as features and inspecting highly-weighted features, we can build domain-specific sentiment lexicons.

To classify postgame tweets in a particular week $k$ in 2012, we train a logistic regression classifier on all postgame tweets starting from 2010 up to but not including week $k$ in 2012. 
We use simple bag-of-words features, conjoining unigrams with an indicator representing whether the tweet is for a home or away team. 
In order to avoid noise from rare unigrams, we only used a unigram feature for a tweet if the unigram appeared in at least 10 tweets during the week that the tweet was written.
We achieved an average accuracy of 67\% over the tested weeks. 
Notable features that were among the top or bottom 30 weighted features are listed in Tab.~\ref{tab:unigramfeats}. Most are intuitive (``win'', ``Great'', etc.). Additionally, we surmise that fans are more likely to comment on the referees (``\awaytag: refs'') after their team loses than after a win.

\begin{small}
\begin{table}[t]
\begin{center}\small
\caption{Highly weighted features for postgame tweet classification. \hometag/\awaytag\ indicates that the unigram is in the tweet for the home or away team, respectively.}\label{tab:unigramfeats}
\begin{tabular} {p{2cm}p{2.25cm}p{1.6cm}|p{1.95cm}p{2.3cm}p{1.5cm}}
\multicolumn{3}{c|}{predicting home team won} & \multicolumn{3}{|c}{predicting away team won} \\ \hline
\hometag: win & \hometag: victory & \awaytag: loss & 
\awaytag: win & \awaytag: congrats & \hometag: lost \\
\hometag: won & \hometag: WIN & \awaytag: lost & 
\awaytag: won & \awaytag: Go & \hometag: loss \\
\hometag: Great & \awaytag: lose & \awaytag: refs & 
\awaytag: Great & \awaytag: proud & \hometag: bad \\
\end{tabular}
\end{center}
\end{table}
\end{small}

\section{Forecasting}
\label{sec:forecasting}
We consider the idea that fan behavior in aggregate can capture meaningful information about upcoming games, and test this claim empirically by using tweets to predict outcomes of NFL games on a weekly basis. We establish baselines using features derived from statistical game data, building upon prior work~\cite{gimpel-06}, and compare accuracies to those of our predictions made using Twitter data.

\subsection{Modeling and Training}
\label{sec:model}
We use a logistic regression classifier to predict game and betting outcomes. 
In order to measure the performance of our feature sets, and tune hyperparameters for our model as the season progresses, we use the following scheme: 
to make predictions for games taking place on week $k \in [4,16]$ in 2012, we use all games from weeks $[1,16]$ of seasons 2010 and 2011, as well as games from weeks $[1,k-3]$ in 2012 as training data.\footnote{We never test on weeks 1--3, and we do not train or test on week 17; it is difficult to predict the earliest games of the season due to lack of historical data and week 17 sees many atypical games among teams that have been confirmed or eliminated from play-off contention.} We then determine the $L_1$ or $L_2$ regularization coefficient from the set $\{0,1,5,10,25,50,100,250,500,1000\}$ that maximizes accuracy on the development set, which consists of weeks $[k-2,k-1]$ of 2012. We follow this procedure to find the best regularization coefficients separately for each feature set and each test week $k$. 
We use the resulting values for final testing on week $k$. We repeat for all test weeks $k\in[4,16]$ in 2012. To evaluate, we compute the accuracy of our predictions across all games in all test weeks. 
We note that these predictions occur in a strictly online setting, and do not consider any information from the future.




\subsection{Features}
\label{sec:features}

\subsubsection{Statistical Game Features} We start with the 10 feature sets shown in Tab.~\ref{tab:featuretable} which only use game statistical data. We began with features from Gimpel~\cite{gimpel-06} and settled upon the feature sets in the table by testing on seasons 2010--2011 using a scheme similar to the one described above. 
These 10 feature sets and the collection of their pairwise unions, a total of 55 feature sets, serve as a baseline to compare to our feature sets that use Twitter data. 

\begin{small}
\begin{table}[t]\small
\caption{List of preliminary feature sets using only game statistics, numbered for reference as $F_i$. $^\ast$Denotes that the features appear for both the home and away teams.}\label{tab:featuretable}
\begin{center}
\begin{tabular} {|p{5.8cm}|p{6.4cm}|}
\hline
point spread line ($F_1$) & over/under line ($F_2$)\\ \hline
avg. points beaten minus missed spread by in current season$^\ast$ ($F_3$) & avg. points beaten minus missed over/under by in current season$^\ast$ ($F_4$) \\ \hline
avg. points scored in current season$^\ast$ ($F_5$) & avg. points given up in current season$^\ast$ ($F_6$) \\ \hline
avg. total points scored in current season$^\ast$ ($F_7$) & avg. (point spread + points scored) in current season$^\ast$ ($F_8$) \\ \hline
home team win WTS percentage in home games in current season & 
avg. interceptions thrown in current season$^\ast$ 
 avg. fumbles lost in current season$^\ast$ \\ 
away team win WTS percentage in away games in current season ($F_9$) &  avg. times sacked in current season$^\ast$ ($F_{10}$) \\\hline
\end{tabular}
\end{center}
\end{table}
\end{small}

\subsubsection{Twitter Unigram Features}
When using tweets to produce feature sets, we first consider an approach similar to the one used in Sec.~\ref{sec:tweetclassification}. In this case, for a given game, we assign the feature (\hometag/\awaytag, unigram) the value $\log(1+$unigram frequency over all weekly tweets assigned to the \hometag/\awaytag\ team). As a means of noise reduction, we only consider (\hometag/\awaytag, unigram) 
pairs occurring in at least $0.1\%$ of the weekly tweets corresponding to the given game; this can be determined before the game takes place. 
This forms an extremely high-dimensional feature space in contrast to the game statistics features, so we now turn to dimensionality reduction. 


\subsubsection{Dimensionality Reduction}
To combine the above two feature sets, we use \textbf{canonical correlation analysis} (CCA)~\cite{hotelling:1936}. We use CCA to simultaneously perform dimensionality reduction on the unigram features and the game statistical features to yield a low-dimensional representation of the total feature space. 

For a paired sample of vectors $\mathbf{x}^i_1 \in \mathbb{R}^{m_1}$ and $\mathbf{x}^i_2 \in \mathbb{R}^{m_2}$, CCA finds a pair of linear transformations of the vectors onto $\mathbb{R}^k$ so as to maximize the correlation of the projected components and so that the correlation matrix between the variables in the canonical projection space is diagonal. While developed to compute the degree of correlation between two sets of variables, it is a good fit for \textbf{multi-view learning problems} in which the predictors can be partitioned into disjoint sets (`views') and each is assumed sufficient for making predictions. Previous work has focused on the semi-supervised setting in which linear transformations are learned from collections of predictors and then regression is carried out on the low dimensional projection, leading to lower sample complexity~\cite{kakade:2007}. Here, we retain the fully supervised setting, but use CCA for dimensionality reduction of our extremely high-dimensional Twitter features. We experiment with several values for the number of components of the reduced matrices resulting from CCA. 
 
\subsubsection{Twitter Rate Features}
As another way to get a lower-dimensional set of Twitter features, we consider a feature that holds a signed representation of the level of increase/decrease in a team's weekly tweet volume compared to the previous week. In computing these \textbf{rate} features, we begin by taking the difference of a team's weekly tweet volume for the week to be predicted $v_{\mathrm{curr}}$, and the team's weekly tweet volume for the previous week in which they had a game $v_{\mathrm{prev}}$ or the team's average weekly tweet volume after its previous game $v_{\mathrm{prevavg}}$. We will use $v_{\mathrm{old}}$ to refer to the subtracted quantity in the difference, either $v_{\mathrm{prev}}$ or $v_{\mathrm{prevavg}}$. 
This difference is mapped to a categorical variable based on the value of a parameter $\Delta$ which determines how significant we consider an increase in volume from $v_{\mathrm{old}}$ to be. Formally, we define a function $\mathrm{rate}_{S}:\mathbb{Z}\times\mathbb{Z}\times\mathbb{N}\rightarrow\{-2,-1,0,1,2\}$, $(v_{\mathrm{old}},v_{\mathrm{curr}},\Delta)\mapsto \mathrm{sign}(v_{\mathrm{curr}}-v_{\mathrm{old}})\lfloor\frac{|v_{\mathrm{curr}}-v_{\mathrm{old}}|}{\Delta}\rfloor$
 that is decreasing in its first argument, increasing in its second argument, and whose absolute value is decreasing in its third argument.

\begin{small}
\begin{table}[t]
\caption{Example of how the $\mathrm{rate}_{S}$ feature is defined with $\Delta =500$ (left) and how the $\mathrm{rate}_{P}$ feature is defined with $\theta =.2$.}\label{tab:rateexamples}
\begin{center}
\begin{tabular} {c | c | c}
$v_{\mathrm{old}}$ & $v_{\mathrm{curr}}$ & $\mathrm{rate}_{S}(v_{\mathrm{old}},v_{\mathrm{curr}},500)$ \\ \hline
2000 & $(3000,\infty)$ & 2 \\ \hline
2000 & $(2500,3000]$ & 1 \\ \hline
2000 & $[1500,2500]$ & 0 \\ \hline
2000 & $[1000,1500)$ & -1 \\ \hline
2000 & $[0,1000)$ & -2
\end{tabular}
\quad
\begin{tabular} {c | c | c}
$v_{\mathrm{old}}$ & $v_{\mathrm{curr}}$ & $\mathrm{rate}_{P}(v_{\mathrm{old}},v_{\mathrm{curr}},.2)$ \\ \hline
2000 & $(2800,\infty)$ & 2 \\ \hline
2000 & $(2400,2800]$ & 1 \\ \hline
2000 & $[1600,2400]$ & 0 \\ \hline
2000 & $[1200,1600)$ & -1 \\ \hline
2000 & $[0,1200)$ & -2
\end{tabular}
\end{center}
\end{table}
\end{small}

This idea of measuring the rate of change in tweet volumes is further generalized by categorizing the difference in volume $(v_{\mathrm{curr}}-v_{\mathrm{old}})$ by computing its percentage of $v_{\mathrm{old}}$, or formally as a function $\mathrm{rate}_{P}:\mathbb{Z}\times\mathbb{Z}\times(0,1]\rightarrow\{-2,-1,0,1,2\}$, $(v_{\mathrm{old}},v_{\mathrm{curr}},\theta)\mapsto \mathrm{sign}(v_{\mathrm{curr}}-v_{\mathrm{old}})\lfloor\frac{|v_{\mathrm{curr}}-v_{\mathrm{old}}|}{\theta \cdot v_{\mathrm{old}}}\rfloor$ which has the same functional properties as the $\mathrm{rate}_{S}$ function. 
Examples of how the $\mathrm{rate}_{S}$ and $\mathrm{rate}_{P}$ functions are defined are provided in Table~\ref{tab:rateexamples}. 
Thus, we may take $v_{\mathrm{old}} = v_{\mathrm{prev}}$ or $v_{\mathrm{old}} = v_{\mathrm{prevavg}}$, and categorize the difference using a static constant $\Delta$ or a percentage $\theta$ of $v_{\mathrm{old}}$, giving us four different versions of the rate feature.

In preliminary testing on the 2010 and 2011 seasons, we found that the $\mathrm{rate}_{S}$ feature worked best with $v_{\mathrm{old}}=v_{\mathrm{prev}}$ and $\Delta = 500$, so we also use these values in our primary experiments below with $\mathrm{rate}_{S}$. 
For $\mathrm{rate}_{P}$, we experiment with
$\theta\in\{0.1,0.2,0.3,0.4,0.5\}$ and $v_{\mathrm{old}}\in\{v_{\mathrm{prev}},v_{\mathrm{prevavg}}\}$. 

\subsection{Experiments}
We consider three prediction tasks: winner, winner WTS, and over/under. Our primary results are shown in Tab.~\ref{tab:allaccs}. We show results for all three tasks for several individual feature sets. We also experimented with many conjunctions of feature sets; the best results for each task over all feature set conjunctions tested are shown in the final three rows of the table.

The Twitter unigram features alone do poorly on the WTS task (47.6\%), but they improve to above 50\% when combined with the statistical features via CCA. 
Surprisingly, however, the Twitter unigram features alone perform better than most other feature sets on over/under prediction, reaching 54.3\%. This may be worthy of follow-up research. 
On winner WTS prediction, the Twitter $\mathrm{rate}_S$ feature (with $v_{\mathrm{prev}}$ and $\Delta=500$) obtains an accuracy above 55\%, which is above the accuracy needed to be profitable after factoring in the bookmaker's commission. We found these hyperparameter settings ($v_{\mathrm{prev}}$ and $\Delta$) based on preliminary testing on the 2011 season, in which they consistently performed better than other values; the success translates to the 2012 season as well. Interestingly, the Twitter rate features perform better on winner WTS than on straight winner prediction, while most statistical feature sets perform better on winner prediction. We see a similar trend in Tab.~\ref{tab:ratepaccs}, which shows results with Twitter $\mathrm{rate}_P$ features with various values for $\theta$ and $v_{\mathrm{old}}$.


We observed in preliminary experiments on the 2011 season that feature sets with high predictive accuracy early on in the season will not always be effective later, necessitating the use of different feature sets throughout the season. For each week $k\in[5,16]$, we use the training and testing scheme described in Sec.~\ref{sec:model} to compute the feature set that achieved the highest accuracy on average over the previous two weeks, starting with week 3. This method of feature selection is similar to our method of tuning regularization coefficients. Over 12 weeks and 177 games in the 2012 season, this strategy correctly predicted the winner \textbf{63.8\%} of the time, the winner WTS \textbf{52.0\%} of the time, and the over under \textbf{44.1\%} of the time. This is a simple way of selecting features and future work might experiment with more sophisticated online feature selection techniques. We expect there to be room for improvement due to the low accuracy on the over/under task (44.1\%) despite there being several feature sets with much higher accuracies, as can be seen in Tab.~\ref{tab:allaccs}. 

\begin{figure}
\begin{center}
\epsfysize=7cm \epsfbox{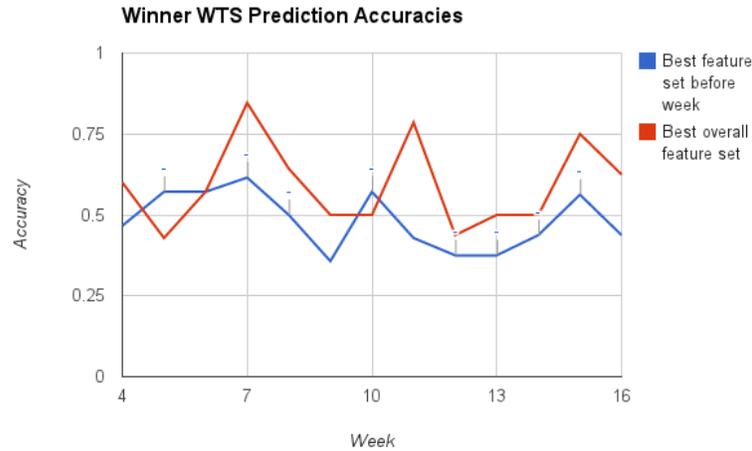}
\caption[]{Weekly accuracies for the best overall feature set in hindsight, and the best feature set leading up to the given week for winner WTS prediction. Marks above the `Best feature set before week' line indicate weeks where the best feature set changed.}\label{fig:chosenfeatplot}
\end{center}
\end{figure}

Another simple method of selecting a feature set for week $k\in[4,16]$ is choosing the feature set achieving the highest accuracy on average over \emph{all} previous weeks, starting with week 3, using the same scheme described in Sec.~\ref{sec:model}. This feature set can be thought of as the best feature set at the point in the season at which it is chosen. In Fig.~\ref{fig:chosenfeatplot} we observe that the best feature set changes very frequently, going through 8 different feature sets in a 13-week period. 

\begin{small}
\begin{table}
\caption{$\mathrm{rate}_{P}$ winner and winner WTS accuracies for different values of $\theta$ and $v_{\mathrm{old}}$.}\label{tab:ratepaccs}
\begin{center}
\begin{tabular} {c|c|c|c|c}
\multicolumn{1}{c}{} & \multicolumn{2}{|c|}{$v_{\mathrm{prev}}$} & \multicolumn{2}{c}{$v_{\mathrm{prevavg}}$} \\ 
$\theta$ & \,\,winner\,\, & \,\,WTS\,\, & \,\,winner\,\, & \,\,WTS\,\, \\\hline
0.1 & 51.0 & 51.4 & 51.0 & 50.0 \\ 
0.2 & 53.8 & 51.0 & 52.4 & 45.7 \\ 
0.3 & 51.4 & 52.4 & 52.4 & 54.3 \\ 
0.4 & 54.8 & 49.5 & 51.4 & 49.5 \\ 
0.5 & 52.9 & 45.2 & 53.4 & 49.5 \\
\end{tabular}
\end{center}
\end{table}
\end{small}



\begin{small}
\begin{table}
\caption{Accuracies across prediction tasks and feature sets. Lower pane shows oracle feature sets for each task, with the highest accuracies starred.}\label{tab:allaccs}
\begin{center}
\begin{tabular} {c | c | c | c}
& \multicolumn{3}{c}{prediction tasks} \\
features & \,\,winner\,\, & \,\,WTS\,\, & \,\,over/under\,\, \\ \hline
point spread line ($F_{1}$) & 60.6 & 47.6 & 48.6 \\ 
over/under line ($F_{2}$) & 52.3 & 49.0 & 48.6 \\ 
$F_{3}$ & 56.3 & 50.0 & 50.0 \\ 
$F_{4}$ & 52.3 & 54.8 & 50.5 \\ 
$F_{5}$ & 65.9 & 51.0 & 44.7 \\ 
$F_{10}$ & 56.7 & 51.4 & 46.6 \\ 
$\bigcup_i F_{i}$ & 63.0 & 47.6 & 51.0 \\ \hline
Twitter unigrams & 52.3 & 47.6  & 54.3 \\ \hline
CCA: $\bigcup_i F_{i}$ and Twitter unigrams, 1 component\, & 47.6 & 50.4 & 43.8 \\ 
\phantom{CCA: $\cup F_{i}$ and Twitter unigrams, } 2 components\, & 47.6 & 51.0 & 43.8 \\ 
\phantom{CCA: $\cup F_{i}$ and Twitter unigrams, } 4 components\, & 50.5 & 51.9 & 44.2 \\ 
\phantom{CCA: $\cup F_{i}$ and Twitter unigrams, } 8 components\, & 47.6 & 48.1 & 42.3 \\ \hline
Twitter $\mathrm{rate}_{S} (v_{\mathrm{prev}},\Delta=500)$ & 51.0 & 55.3 & 52.4 \\ \hline \hline
$F_{5}\cup F_{9}\cup$ Twitter  $\mathrm{rate}_{P} (v_{\mathrm{prev}},\theta =.2)$ & 65.9$^\ast$ & 51.4 & 48.1 \\ 
$F_{3}\cup F_{10}\cup$ Twitter $\mathrm{rate}_{P} (v_{\mathrm{prev}},\theta =.1)$ & 56.3 & 57.2$^\ast$ & 48.1 \\ 
$F_{3}\cup F_{4}\cup$ Twitter  $\mathrm{rate}_{S} (v_{\mathrm{prev}},\Delta =200)$ & 54.8 & 49.0 & 58.2$^\ast$
\end{tabular}
\end{center}
\end{table}
\end{small}

\section{Conclusion}

We introduced a new dataset that includes a large volume of tweets aligned to NFL games from the 2010--2012 seasons. We explored a range of feature sets for predicting game outcomes, finding that simple feature sets that use Twitter data could match or exceed the performance of game statistics features. Our dataset is made available for academic research at \url{www.ark.cs.cmu.edu/football}.

\begin{small}
\subsubsection{Acknowledgments} We thank the anonymous reviewers, Scott Gimpel at NFLdata.com, Brendan O'Connor, Bryan Routledge, and members of the ARK research group. This research was supported in part by the National Science Foundation (IIS-1054319) and Sandia National Laboratories (fellowship to K. Gimpel). 
\end{small}

\bibliographystyle{splncs03}
\bibliography{mybiblio}
\end{document}